\documentclass[11pt]{amsart}

\usepackage{amsmath,amssymb,amsfonts,mathrsfs}
\usepackage{hyperref}
\usepackage{geometry}
\theoremstyle{definition}

\newcommand{\fg}{\mathfrak{g}}

\title{From Yang--Mills to Yang--Baxter:\\
In Memory of Rodney Baxter and Chen--Ning Yang}

\author{Bai-Ling Wang}
\address{Mathematical Sciences Institute,\\
Australian National University}
\email{bai-ling.wang@anu.edu.au}

\date{}


\begin{document}

\begin{abstract}
The year 2025 marked the passing of two towering figures of twentieth-century mathematical physics, Rodney Baxter and Chen–Ning Yang. Yang reshaped modern physics through the introduction of non-abelian gauge theory and, independently, through the consistency conditions underlying what is now called the Yang–Baxter equation. Baxter transformed those conditions into a systematic theory of exact solvability in statistical mechanics and quantum integrable systems.

This article is written in memory of Baxter and Yang, whose work revealed how local consistency principles generate global mathematical structure. We review the Yang–Mills formulation of gauge theory, its mass obstruction and resolution via symmetry breaking, and the geometric framework it engendered, including instantons, Donaldson-Floer theory, magnetic monopoles, and Hitchin systems. In parallel, we trace the emergence of the Yang–Baxter equation from factorised scattering to solvable lattice models, quantum groups, and Chern–Simons theory. Rather than separate narratives, gauge theory and integrability are presented as complementary manifestations of a shared coherence principle -- an ongoing journey from gauge symmetry toward mathematical unity.
\end{abstract}

\maketitle

%\tableofcontents
%================================================
%================================================
\section{Introduction: Two Lives, One Mathematical Legacy}
%================================================

The year 2025 marked the passing of two of the most influential architects of
modern theoretical physics, Rodney Baxter and Chen--Ning Yang.
Their scientific lives unfolded along different paths, one rooted in quantum
field theory and fundamental symmetries, the other in statistical mechanics and
exact solvability, yet their ideas converged in a profound and enduring way.
Few pairs of scientists illustrate so vividly how deep physical insight can
transcend disciplinary boundaries and reshape both physics and mathematics.

Chen--Ning Yang belonged to a generation for whom symmetry was not merely a
technical tool, but a guiding philosophical principle.
Trained in the aftermath of the Second World War, Yang approached physics with
a rare blend of conceptual audacity and mathematical clarity.
His early work already displayed a defining trait: an insistence that physical
laws should be formulated in the most intrinsic and universal language
available.
This viewpoint culminated in the 1954 collaboration \cite{YangMills1954} with Robert Mills, where
Yang made a leap that was, at the time, almost shocking, internal symmetries,
like isotopic spin, should be treated on the same footing as spacetime
symmetries, and should therefore be local.

The Yang--Mills paper did not solve an outstanding experimental puzzle, nor did
it immediately explain known data.
Instead, it proposed a new \emph{way of thinking} about interactions.
Yang later remarked that the work was motivated less by phenomenology than by
consistency and elegance.
In retrospect, this was a decisive moment in theoretical physics: a theory
whose physical relevance was initially uncertain would later become the
structural backbone of the Standard Model and a wellspring of modern geometry.

Rodney Baxter’s scientific journey \cite{Baxter17} was shaped by a different, but equally
demanding, intellectual challenge: how to extract exact results from strongly
interacting many-body systems.
In an era when approximate and numerical methods dominated statistical
mechanics, Baxter pursued the almost heretical idea that certain models could
be solved \emph{exactly}, not accidentally but systematically.
His work revealed that solvability is governed by rigid algebraic constraints,
and that once these constraints are identified, entire classes of models fall
into place.

One of Baxter’s most remarkable achievements was to recognise the unifying role
of an equation that had appeared, somewhat quietly, in the work of Yang on
one-dimensional scattering.
What had begun as a consistency condition ensuring that different sequences of
pairwise scatterings lead to the same outcome was elevated by Baxter into a
central organising principle.
The Yang--Baxter equation became the cornerstone of integrable systems,
governing transfer matrices, commuting operators, and ultimately the exact
calculation of physical quantities.

The intellectual resonance between Yang and Baxter lies not in a shared
subject, but in a shared philosophy.
Both believed that deep structure reveals itself through consistency.
For Yang, local gauge invariance enforces the geometry of interactions.
For Baxter, algebraic consistency enforces exact solvability.
In both cases, local rules determine global behaviour, and simple principles
give rise to unexpectedly rich mathematics.

This shared legacy has had consequences far beyond their original contexts.
Yang--Mills theory evolved into a central pillar of modern mathematics,
influencing topology, geometry, and representation theory.
The Yang--Baxter equation, through quantum groups and braided categories,
reshaped algebra, low-dimensional topology, and mathematical physics.
Today, these two strands meet in areas such as integrable gauge theories,
topological quantum field theory, string theory, and mirror symmetry.

This article is written in memory of Rodney Baxter and Chen--Ning Yang, not as a
chronological biography, but as a reflection on the ideas they set in motion.
Their work teaches us that the most enduring advances often arise not from
answering existing questions, but from asking new ones  that force us
to rethink the language of mathematics in which physical laws are written and studied.
%====================================%================================================
\section{Yang and Mills (1954): Isotopic Gauge Invariance}\label{sec:YM1954}
%================================================

This section gives a technical account of the Yang--Mills construction in the
original $SU(2)$ isotopic-spin setting, phrased in modern geometric language
(while remaining close in spirit to the 1954 presentation).
We work on Minkowski space $\mathbb{R}^{1,3}$ with coordinates $x^\mu$
($\mu=0,1,2,3$) and metric $\eta=\mathrm{diag}(1,-1,-1,-1)$; repeated indices
are summed.

Before introducing gauge fields, it is important to recall the Dirac
Lagrangian
\[
\int \mathcal{L}_{\mathrm{Dirac}} d^4 x 
=\int d^4 x\bar\psi\,(i\gamma^\mu\partial_\mu - m)\psi,
\]
which provides the relativistic quantum description of
spin-$\tfrac12$ matter. 
The Dirac Lagrangian occupies a central position in both mathematics and
physics as the first fully relativistic field theory describing spin-$\tfrac12$
matter.
Introduced through Dirac’s 1928 equation, it immediately lessened the tension
between quantum mechanics and special relativity and led to the prediction of
antiparticles, a striking confirmation of the theory’s physical depth.
Its rapid and successful application in quantum electrodynamics established
the Dirac formalism as the universal language for fermionic matter.
From a geometric viewpoint, the Dirac Lagrangian realises the Clifford algebra
of spacetime as a first-order differential operator acting on the spinors, encoding Lorentz symmetry and the causal structure of spacetime in an
intrinsic way.
Its invariance under global internal symmetries governs how matter fields may
transform, and its success in quantum electrodynamics demonstrated that
interactions can be introduced consistently by replacing ordinary derivatives
with covariant ones.
This replacement principle is the precise mechanism by which gauge connections
enter the theory, and it is this structural feature that Yang and Mills
generalised to non-abelian internal symmetries.
We briefly explain the notation appearing in the Dirac Lagrangian density.

\medskip
\noindent\textbf{The spinor field $\psi$.}
The field $\psi(x)$ is a \emph{Dirac spinor}, i.e. a complex-valued function on
spacetime taking values in $\mathbb{C}^4$:
\[
\psi : \mathbb{R}^{1,3} \longrightarrow \mathbb{C}^4.
\]
It represents a relativistic fermionic particle (such as a nucleon or electron)
with spin $\tfrac12$.
In the Yang--Mills setting, $\psi$ may also carry internal indices (e.g.\ an
$SU(2)$ isospin index), but these are suppressed in the notation.

\medskip
\noindent\textbf{Gamma matrices $\gamma^\mu$.}
The matrices $\gamma^\mu$ ($\mu=0,1,2,3$) are $4\times4$ complex matrices
satisfying the Clifford algebra relations
\[
\{\gamma^\mu,\gamma^\nu\}
:= \gamma^\mu\gamma^\nu+\gamma^\nu\gamma^\mu
=2\eta^{\mu\nu}I_4,
\]
where $\eta^{\mu\nu}=\mathrm{diag}(1,-1,-1,-1)$ is the Minkowski metric and $I_4$
is the $4\times4$ identity matrix.
They encode the spinorial representation of the Lorentz group and ensure
relativistic covariance of the theory.

A concrete realization of the Clifford algebra
is provided by the \emph{Dirac   representation}.
In this representation the gamma matrices are written in block form using the
$2\times2$ identity matrix $I_2$ and the Pauli matrices
\[
\sigma^1=
\begin{pmatrix}
0 & 1\\
1 & 0
\end{pmatrix},\qquad
\sigma^2=
\begin{pmatrix}
0 & -i\\
i & 0
\end{pmatrix},\qquad
\sigma^3=
\begin{pmatrix}
1 & 0\\
0 & -1
\end{pmatrix}.
\]

The Dirac gamma matrices are then given explicitly by
\[
\gamma^0 =
\begin{pmatrix}
I_2 & 0 \\
0 & -I_2
\end{pmatrix},
\qquad
\gamma^i =
\begin{pmatrix}
0 & \sigma^i \\
-\sigma^i & 0
\end{pmatrix},
\quad i=1,2,3.
\]

\medskip
\noindent\textbf{The Dirac adjoint $\bar\psi$.}
The symbol
\[
\bar\psi := \psi^\dagger \gamma^0
\]
denotes the \emph{Dirac adjoint} of $\psi$, where $\psi^\dagger$ is the Hermitian
conjugate (complex conjugate transpose) of the column vector $\psi$.
This definition ensures that the scalar quantity $\bar\psi\psi$ is Lorentz
invariant and that the Lagrangian density is real.

\medskip
\noindent\textbf{The derivative $\partial_\mu$.}
The operator
$
\partial_\mu = \frac{\partial}{\partial x^\mu}
$
denotes partial differentiation with respect to the spacetime coordinate
$x^\mu$.

\medskip
\noindent\textbf{The mass parameter $m$.}
The constant $m>0$ is the physical mass of the fermion.
The term $-m\bar\psi\psi$ couples the left- and right-handed components of the
spinor and is responsible for the rest energy of the particle.

\medskip
\noindent\textbf{The imaginary unit $i$.}
The factor $i=\sqrt{-1}$ ensures that the Dirac operator
$i\gamma^\mu\partial_\mu$ is formally self-adjoint with respect to the natural
inner product on spinors, leading to a real action functional and unitary time
evolution.

\medskip
\noindent\textbf{Physical equation of motion.}
The Euler--Lagrange equation associated to $\mathcal{L}_{\mathrm{Dirac}}$ is the
Dirac equation
\[
(i\gamma^\mu\partial_\mu - m)\psi=0,
\]
which is the relativistic wave equation for a free spin-$\tfrac12$ particle.

%------------------------------------------------
\subsection{Global isotopic symmetry and its localisation}\label{subsec:global_to_local}
%------------------------------------------------
 
A fundamental structural feature of 
the free Dirac Lagrangian density.
\begin{equation}\label{eq:dirac_lagrangian}
\mathcal{L}_{\mathrm{Dirac}}
=\bar\psi\,(i\gamma^\mu\partial_\mu - m)\psi,
\end{equation}
is its invariance
under certain global transformations of the spinor field,  the {\em global phase symmetry}
\begin{equation}\label{eq:global_phase}
\psi(x)\longmapsto \psi'(x)=e^{i\alpha}\psi(x),
\qquad \alpha\in\mathbb{R}\ \text{constant}.
\end{equation}

More generally, suppose that the spinor field carries an internal index and
takes values in $\mathbb{C}^n$, so that
\[
\psi(x)\in\mathbb{C}^4\otimes\mathbb{C}^n.
\]
Let $G$ be a compact Lie group acting unitarily on $\mathbb{C}^n$.
For example, in the isotopic spin model considered by Yang and Mills,
$G=SU(2)$ and $n=2$, corresponding to a proton--neutron doublet.

The Dirac Lagrangian density \eqref{eq:dirac_lagrangian} remains invariant under the
\emph{global} internal symmetry
\begin{equation}\label{eq:global_internal}
\psi(x)\longmapsto \psi'(x)=U\psi(x),
\qquad U\in G
\end{equation}
because $U$ commutes with $\gamma^\mu$ and with $\partial_\mu$.
This invariance expresses the conservation of isotopic spin.

Yang and Mills proposed to \emph{localise} the internal symmetry by allowing the
group element to vary with spacetime:
\begin{equation}\label{eq:local_internal}
\psi(x)\longmapsto \psi'(x)=g(x)\psi(x),\  \text{for }\ 
 g:\mathbb{R}^{1,3} \longrightarrow  G.
\end{equation}
While the mass term in \eqref{eq:dirac_lagrangian} remains invariant, the kinetic
term does not.
Indeed,
\[
\bar\psi'(i\gamma^\mu\partial_\mu)\psi'
=\bar\psi i\gamma^\mu\bigl((g^{-1}\partial_\mu g)\psi+ \partial_\mu\psi\bigr),
\]
which contains additional terms involving $g^{-1}\partial_\mu g$.
Thus the ordinary derivative $\partial_\mu$ is incompatible with local internal
symmetry.

The failure of invariance under \eqref{eq:local_internal} is not an inconsistency
but a signal.
It indicates that the derivative must be modified in a way that reflects the
geometry of a vector bundle with structure group $G$.
The required modification is the introduction of a covariant derivative defined
by a connection, which transforms so as to compensate for the spacetime
dependence of $g(x)$.
This necessity is the starting point of Yang--Mills theory, 
as in electromagnetism, local symmetry forces a new field: \emph{the non-abelian gauge field}.

%------------------------------------------------
\subsection{Gauge potential as a connection}\label{subsec:connection}
%------------------------------------------------

Yang and Mills introduced an $\mathfrak{su}(2)$-valued gauge potential (an local connection 1-form)
\[
A(x) =A_\mu (x) \,dx^\mu,\qquad A_\mu(x)\in\mathfrak{su}(2).
\]
Choose a basis $T^a$ ($a=1,2,3$) of $\mathfrak{su}(2)$ with
\[
[T^a,T^b]=\epsilon^{abc}T^c,\qquad \mathrm{tr}(T^aT^b)=-\frac12\delta^{ab},
\]
then $A_\mu=A_\mu^aT^a$. 
They defined the covariant derivative on the doublet by
\[
D_\mu\psi:=\partial_\mu\psi+ A_\mu\psi.
\]

The local gauge transformation $g(x) \in SU(2)$ acts by
\[
\psi \mapsto \psi' = g\psi,\qquad
A_\mu \mapsto A_\mu' = gA_\mu g^{-1} -  (\partial_\mu g)\,g^{-1}.
\]
A direct computation verifies the key covariance property:
\[
D_\mu\psi' = g\,D_\mu\psi.
\]
Thus any Lagrangian built from $\bar\psi\gamma^\mu D_\mu\psi$ will be gauge-invariant.
This is precisely the transformation law of a connection on a principal $SU(2)$-bundle.
On $\mathbb{R}^{1,3}$ the bundle is topologically trivial, but the formalism
extends immediately to general manifolds and nontrivial bundles—one of the
routes by which Yang-Mills theory entered modern geometry.

%------------------------------------------------
\subsection{Curvature and the non-abelian field strength}\label{subsec:curvature}
%------------------------------------------------

The field strength of the gauge field is  a (globally defined) $\mathfrak{su}$-valued 2-form,  the curvature of the connection:
\[
F:=dA+A\wedge A,
\qquad\text{so that}\qquad
F=\frac12\,F_{\mu\nu}\,dx^\mu\wedge dx^\nu,
\]
with components
\[
F_{\mu\nu}
= \partial_\mu A_\nu-\partial_\nu A_\mu + [A_\mu,A_\nu].
\]

Two fundamental structural features already appear here:

\begin{itemize}
\item \emph{Nonlinearity / self-interaction.}
The commutator term $[A_\mu,A_\nu]$ implies the gauge field interacts with itself,
in contrast to Maxwell theory (the abelian case $U(1)$, where the commutator vanishes).

\item \emph{Gauge covariance of curvature.}
Under a gauge transformation,
\[
F_{\mu\nu}' = g F_{\mu\nu} g^{-1}.
\]
Hence any expression formed from an adjoint-invariant bilinear form on $\mathfrak{su}(2)$,
such as $\mathrm{tr}(F_{\mu\nu}F^{\mu\nu})$, is gauge-invariant.
\end{itemize}

The curvature satisfies the (non-abelian) Bianchi identity:
\[
D_A F = 0,
\quad\text{i.e.}\quad
D_{[\lambda}F_{\mu\nu]}=0,
\]
where $D_A$ is the covariant exterior derivative induced by $A$.

%------------------------------------------------
\subsection{Yang-Mills action and Euler-Lagrange equations}\label{subsec:ym_action}
%------------------------------------------------

The Yang-Mills Lagrangian density is
\[
\mathcal{L}_{\mathrm{YM}}
= -\frac{1}{2}\,\mathrm{tr}(F_{\mu\nu}F^{\mu\nu}).
\]
The corresponding action is
\[
S_{\mathrm{YM}}[A]=\int\mathcal{L}_{\mathrm{YM}}\,d^4x.
\]
The critical points of $S_{\mathrm{YM}}$ is given by the Euler-Lagramge equation of the Yang-Mills action, the most celebrated  Yang-Mills equation
\[
D_A^{*}F=  D^\mu F_{\mu\nu}=0.
\]

\medskip
\noindent\textbf{Comparison with Maxwell.}
For $U(1)$, $A_\mu$ is scalar-valued, $F_{\mu\nu}=\partial_\mu A_\nu-\partial_\nu A_\mu$,
and $D^\mu F_{\mu\nu}=\partial^\mu F_{\mu\nu}=0$ recovers the source-free Maxwell equations.
Thus Yang--Mills is the canonical non-abelian generalisation.

%------------------------------------------------%------------------------------------------------
\subsection{Coupling to matter, gauge invariance, and gauge fixing}
\label{subsec:matter_coupling}
%------------------------------------------------

Let $\psi$ be a Dirac spinor transforming in a unitary representation of a
compact Lie group $G$, and let $A_\mu$ be a $\mathfrak{g}$-valued gauge
connection.
The gauge-invariant matter Lagrangian density is
\begin{equation}\label{eq:matter_lagrangian}
\mathcal{L}_{\mathrm{matter}}
=\bar\psi\,(i\gamma^\mu D_\mu - m)\psi,
\qquad
D_\mu= \partial_\mu + A_\mu .
\end{equation}
The full Yang--Mills--Dirac Lagrangian density is
\begin{equation}\label{eq:YM_Dirac_Lagrangian}
\mathcal{L}
= -\frac12 \mathrm{tr}(F_{\mu\nu}F^{\mu\nu})
+ \bar\psi\,(i\gamma^\mu D_\mu - m)\psi .
\end{equation}

%------------------------------------------------
\subsubsection*{Gauge invariance of the coupled equations}
%------------------------------------------------

Under a local gauge transformation
\[
\psi \longmapsto \psi' = g(x)\psi,
\qquad
A_\mu \longmapsto A_\mu'
= gA_\mu g^{-1} - (\partial_\mu g)g^{-1},
\]
the covariant derivative transforms equivariantly:
\[
D_\mu \psi \longmapsto D_\mu'\psi'
= g(x)\,D_\mu\psi .
\]
Consequently, the Dirac operator satisfies
\[
(i\gamma^\mu D_\mu' - m)\psi'
= g(x)(i\gamma^\mu D_\mu - m)\psi,
\]
and the Dirac equation
\begin{equation}\label{eq:dirac_gauge}
(i\gamma^\mu D_\mu - m)\psi = 0
\end{equation}
is \emph{gauge invariant}.
Similarly, the curvature transforms by conjugation,
\[
F_{\mu\nu}' = g F_{\mu\nu} g^{-1},
\]
which implies invariance of the Yang--Mills action and of the Yang--Mills
equation
\begin{equation}\label{eq:YM_gauge}
D^\mu F_{\mu\nu} = J_\nu .
\end{equation}

%------------------------------------------------
\subsubsection*{Gauge current and covariant conservation}
%------------------------------------------------

Varying the action with respect to $A_\mu$ yields the matter current
\begin{equation}\label{eq:gauge_current}
J_\nu = \bar\psi \gamma_\nu T^a \psi \, T^a ,
\end{equation}
where $\{T^a\}$ is a basis of $\mathfrak{su}(2)$.
Under a gauge transformation,
\[
J_\nu \longmapsto J_\nu' = g J_\nu g^{-1},
\]
so the current transforms in the adjoint representation.

Taking the covariant divergence of \eqref{eq:YM_gauge} and using the Bianchi
identity, we obtain
\begin{equation}\label{eq:covariant_conservation}
D^\nu J_\nu = 0 .
\end{equation}
This is the non-abelian generalisation of charge conservation.
Unlike the abelian case, $\partial^\nu J_\nu=0$ does \emph{not} hold in general;
only covariant conservation is meaningful.

%------------------------------------------------
\subsubsection*{Gauge redundancy and the need for gauge fixing}
%------------------------------------------------

The equations \eqref{eq:dirac_gauge} and \eqref{eq:YM_gauge} are invariant under
an infinite-dimensional group of local gauge transformations.
As a consequence, the gauge potential $A_\mu$ is not uniquely determined by the
physical fields.
This redundancy is geometric: different connections related by gauge
transformations describe the same physical configuration.

To obtain a well-posed classical initial value problem, or to define a quantum
theory via functional integration, one must impose a \emph{gauge fixing
condition} that selects a representative from each gauge orbit.

%------------------------------------------------
\subsubsection*{Common gauge fixing conditions}
%------------------------------------------------

Several gauge choices are commonly used:

\begin{itemize}
\item \textbf{Lorenz gauge}:
\[
\partial^\mu A_\mu = 0 .
\]
This condition is Lorentz invariant and leads to hyperbolic field equations,
making it particularly suitable for relativistic analysis.

\item \textbf{Temporal gauge}:
\[
A_0 = 0 .
\]
This choice is natural in Hamiltonian formulations and highlights the role of
spatial connections and Gauss-law constraints.

\item \textbf{Coulomb gauge}:
\[
D^\mu A_\mu = 0 .
\]
This gauge is useful in canonical quantisation but breaks manifest Lorentz
invariance.
\end{itemize}

Gauge fixing does not break gauge symmetry at the level of physical
observables; it merely removes redundancy in the description of gauge fields.
In geometric terms, it corresponds to choosing a local slice for the action of the gauge transformation group.

%------------------------------------------------
\section{Spontaneous symmetry breaking and the Higgs mechanism}
%------------------------------------------------

Yang and Mills observed that local non-abelian gauge invariance uniquely fixes
the form of the interaction.
However, it simultaneously forbids the inclusion of a mass term for the gauge
field.
An explicit term of the form $m^2 A_\mu A^\mu$ violates gauge invariance and
destroys the geometric interpretation of $A_\mu$ as a connection.
In modern geometric language, gauge fields are connections on principal
$G$-bundles, and gauge transformations act as bundle automorphisms.
A mass term would break this equivalence by selecting preferred representatives
within gauge orbits.
Thus, the mass obstruction reflects the rigidity of the gauge principle rather
than a failure of the theory.

For nearly a decade, this rigidity was widely viewed as fatal for non-abelian
gauge theories as models of short-range forces.
The decisive conceptual advance came in the mid-1960s with the independent work
of Brout-Englert (\cite{BroEng1964})
and of Higgs (\cite{Higgs1964}). 
They showed that gauge symmetry need not be manifest in the vacuum state.
A theory may possess an exact local symmetry at the level of its Lagrangian,
while the vacuum selects a non-invariant configuration.

The Higgs mechanism provides an explicit realisation of this idea.
A scalar field with a symmetry-breaking potential defines a vacuum manifold,
and choosing a vacuum reduces the symmetry group from $G$ to a subgroup $H$.
The covariant derivative couples the scalar field to the gauge field, and when
expanded around the vacuum, generates quadratic terms for components of the
gauge field in $\mathfrak{g}/\mathfrak{h}$.

Crucially, these mass terms arise from the geometry of the covariant derivative,
not from explicit symmetry breaking.
Gauge invariance remains exact, and the theory retains its geometric meaning as
a theory of connections on a reduced bundle.

%------------------------------------------------
\subsection{Mathematical formulation of the Higgs mechanism}
\label{subsec:higgs_mechanism_math}
%------------------------------------------------

We give a precise geometric and analytic formulation of the Higgs mechanism,
emphasising its interpretation in terms of bundle reduction and elliptic
operators.

\subsubsection*{Gauge fields and Higgs fields}

Let $G$ be a compact Lie group with Lie algebra $\mathfrak g$, and let
$P \to M$ be a principal $G$-bundle over a Riemannian manifold $(M,g)$, which we
interpret as spacetime.
A Yang--Mills field is a connection
\[
A \in \Omega^1(M,\mathfrak g_P),
\]
where $\mathfrak g_P := P\times_G \mathfrak g$ is the adjoint bundle.
Its curvature is
\[
F_A = d A + A\wedge A \in \Omega^2(M,\mathfrak g_P).
\]

Let $(V,\rho)$ be a finite-dimensional unitary representation of $G$, and let
\[
E := P\times_G V
\]
be the associated Hermitian vector bundle.
A Higgs field is a smooth section
\[
\Phi \in \Gamma(E).
\]
The covariant derivative induced by $A$ is
\[
D_A \Phi := d\Phi + \rho_*(A)\Phi \in \Omega^1(M,E).
\]

\subsubsection*{Yang--Mills--Higgs functional}

Fix constants $v>0$ and $\lambda>0$.
The Yang--Mills--Higgs energy functional is
\begin{equation}\label{eq:YMH_functional}
\mathcal{E}(A,\Phi)
=
\int_M
\left(
\frac12 \|F_A\|^2
+
\frac12 \|D_A \Phi\|^2
+
\frac{\lambda}{4}\bigl(\|\Phi\|^2 - v^2\bigr)^2
\right)
\,\mathrm{vol}_g.
\end{equation}
This functional is invariant under the gauge group
$\mathcal{G} = \mathrm{Aut}_G(P)$ acting by
\[
(A,\Phi)\mapsto (gAg^{-1}-dg\,g^{-1},\; \rho(g)\Phi).
\]

\subsubsection*{Vacuum manifold and reduction of structure group}

The vacuum configurations are the absolute minima of the potential term,
characterised by
\[
\|\Phi\| = v.
\]
Pointwise, the set of minima is a homogeneous space
\[
\mathcal V \cong G/H,
\]
where $H\subset G$ is the stabiliser of a chosen vacuum vector
$\Phi_0\in V$.
A choice of vacuum defines a section of the associated bundle
\[
P\times_G(G/H),
\]
and hence a reduction of the structure group of $P$ from $G$ to $H$,
yielding a principal $H$-subbundle
\[
P_H \subset P.
\]

Thus, spontaneous symmetry breaking is mathematically realised as a reduction
of structure group rather than a loss of gauge symmetry.

\subsubsection*{Mass operator for gauge fields}

Decompose the Lie algebra as an $\mathrm{Ad}(H)$-invariant orthogonal sum
\[
\mathfrak g = \mathfrak h \oplus \mathfrak m,
\]
where $\mathfrak h = \mathrm{Lie}(H)$.
Correspondingly, the adjoint bundle splits as
\[
\mathfrak g_P = \mathfrak h_P \oplus \mathfrak m_P.
\]

Consider the quadratic part of the functional \eqref{eq:YMH_functional} in the
gauge field, expanded around a vacuum configuration $(A_0,\Phi_0)$ with
$D_{A_0}\Phi_0=0$.
The term
\[
\|D_A \Phi_0\|^2
=
\|\rho_*(A_{\mathfrak m})\Phi_0\|^2
\]
defines a positive semidefinite bundle endomorphism
\[
\mathcal{M} : \mathfrak m_P \to \mathfrak m_P,
\qquad
\mathcal{M}(X) = \rho_*(X)\Phi_0,
\]
called the \emph{mass operator}.
Its eigenvalues determine the squared masses of gauge bosons associated with
broken generators.
Sections of $\mathfrak h_P$ lie in the kernel of $\mathcal{M}$ and correspond to
massless gauge fields.

From a unifying perspective, the Higgs mechanism plays a dual role.
Locally, it endows gauge bosons with mass by reducing gauge symmetry.
Globally, it enables finite-energy solitons whose masses are fixed by topology
and the symmetry-breaking scale.
This dual role will reappear in later developments, including electric--magnetic
duality, Seiberg--Witten theory, and mirror symmetry.

%------------------------------------------------
\subsection{The ’t Hooft-Polyakov monopole \cite{tHooft, Polyakov}: an example}
\label{subsec:tHooft_Polyakov_monopole_math}
%------------------------------------------------
In this subsection, we describe the classical ’t Hooft-Polyakov monopole as a smooth, finite-energy
solution of the Yang-Mills-Higgs equations in three dimensions.
Throughout, let $G=SU(2)$ (or $SO(3)$) with Lie algebra $\mathfrak g=\mathfrak{su}(2)$,
and let $\langle\cdot,\cdot\rangle$ be an $\mathrm{Ad}$-invariant inner product
on $\mathfrak g$.

\subsubsection*{Fields and energy functional}

Let $A$ be a connection on the trivial principal $G$-bundle over $\mathbb{R}^3$,
identified with a $\mathfrak g$-valued 1-form, and let
\[
F_A = dA +   A\wedge A
\]
be its curvature.
Let $\Phi:\mathbb{R}^3\to \mathfrak g$ be an adjoint Higgs field.
The covariant derivative is
\[
D_A\Phi = d\Phi + [A,\Phi].
\]
Fix constants $\lambda>0$ and $v>0$, and consider the symmetry-breaking potential
\[
V(\Phi)=\frac{\lambda}{4}\bigl(\|\Phi\|^2-v^2\bigr)^2.
\]
The (static) Yang--Mills--Higgs energy functional is
\begin{equation}\label{eq:YMH_energy}
E(A,\Phi)
=
\int_{\mathbb{R}^3}
\left(
\frac12 \|F_A\|^2
+
\frac12 \|D_A\Phi\|^2
+
V(\Phi)
\right)\, d^3x.
\end{equation}

\subsubsection*{Topological charge and reduction to $U(1)$}

The \emph{finite-energy configuration}  implies that the \emph{direction}
\[
\hat\Phi := \frac{\Phi}{\|\Phi\|} : S^2_\infty \longrightarrow S^2
\]
is well-defined on the sphere at infinity $S^2_\infty$.
Its degree
\[
k := \deg(\hat\Phi)\in \pi_2(S^2)\cong \mathbb{Z}
\]
is the \emph{magnetic charge}.
Equivalently, the choice of a vacuum direction at infinity reduces the structure
group from $SU(2)$ to the stabiliser $U(1)$, producing an induced principal
$U(1)$-bundle over $S^2_\infty$ whose first Chern class equals $k$.

In the Prasad--Sommerfield (BPS) limit $\lambda\to 0$, the potential term vanishes
and one obtains a sharp energy bound.
Using $\|F_A\mp *D_A\Phi\|^2\ge 0$ and integration by parts, one derives
\begin{equation} E(A,\Phi)
=
\frac12\int_{\mathbb{R}^3}\|F_A- *D_A\Phi\|^2\,d^3x
\ + \ \int_{\mathbb{R}^3}\langle F_A\wedge D_A\Phi\rangle \geq \int_{S^2_\infty} \langle F_{A_\infty}, \hat \Phi\rangle. 
\label{eq:Bogomolny_bound}
\end{equation}
The equality in \eqref{eq:Bogomolny_bound} holds if and only if $(A,\Phi)$ satisfies the \emph{Bogomolny equation} \cite{Bogo1976}. 
\begin{equation}\label{eq:Bogomolny_equation}
F_A = * D_A\Phi.
\end{equation}
Solutions of \eqref{eq:Bogomolny_equation} are called magnetic 
(\emph{BPS}) monopoles when the magnetic charge $k\geq 0$.

The historical arc from Yang and Mills to Higgs, ’t Hooft, and Polyakov reveals a
remarkable unity.
What appeared initially as a fatal obstruction to gauge theory was resolved not
by abandoning the gauge principle, but by understanding it more deeply.
Spontaneous symmetry breaking preserves gauge invariance while reshaping the
vacuum geometry.
The Higgs mechanism generates mass through covariant geometry.
Renormalisability secures quantum consistency.
Magnetic monopoles expose the global and topological content of the theory.
Together, these developments completed Yang--Mills theory as both a physical
and mathematical framework.
They transformed a symmetry principle into a generative engine for geometry,
topology, and nonperturbative phenomena—a role that continues to shape modern
mathematical physics.

%================================================
\section{Yang--Mills Theory and the Mathematical Revolution}
%================================================

\noindent
With the conceptual resolution of the mass problem through symmetry breaking,
Yang--Mills theory became not only physically viable but mathematically fertile.
In particular, when reformulated on general four-manifolds and studied through
its variational and elliptic properties, the Yang--Mills equations revealed a
depth far beyond their original role in particle physics.
This shift from dynamics on spacetime to geometry on manifolds initiated a
mathematical revolution in which gauge fields became fundamental objects of
differential topology, global analysis, and geometry.
We begin with the anti-self-dual Yang--Mills equations in four dimensions and
their role in Donaldson and Floer theory.

%------------------------------------------------
\subsection{Anti-self-dual Yang--Mills equations and Donaldson--Floer theory}
\label{subsec:ASD_Donaldson}
%------------------------------------------------

The most profound mathematical consequences of Yang--Mills theory emerge in
four dimensions.
This is not accidental: only in dimension four does the Hodge star operator act
as an involution on two-forms, allowing the curvature of a connection to be
decomposed into self-dual and anti-self-dual components.
This special feature places four-dimensional gauge theory at the intersection
of elliptic analysis, geometry, and topology.

\subsubsection*{Yang--Mills functional and self-duality}
Let $X$ be a closed, oriented Riemannian four-manifold, and let $P\to X$ be a
principal $G$-bundle with compact structure group $G$ (typically $SU(2)$ or
$SO(3)$).
For a connection $A$ on $P$, the Yang--Mills functional is
\[
\mathrm{YM}(A)
= \int_X \langle F_A \wedge *F_A\rangle,
\]
where $F_A$ is the curvature and $\langle\cdot,\cdot\rangle$ is an invariant
inner product on the Lie algebra.

In four dimensions, the Hodge star induces an orthogonal decomposition
\[
\Omega^2(X,\mathfrak{g})
= \Omega^2_+(X,\mathfrak{g}) \oplus \Omega^2_-(X,\mathfrak{g}),
\]
into self-dual and anti-self-dual two-forms, under which the curvature 
\[
F_A = F_A^+ + F_A^-. 
\]
can be written as the sum of self-dual and anti-self-dual components. 
A standard computation yields the identity
\[
\mathrm{YM}(A)
= \|F_A^+\|^2 + \|F_A^-\|^2
\;\ge\; 2\bigl|\!\int_X \langle F_A\wedge F_A\rangle\bigr|,
\]
with equality if and only if either $F_A^+=0$ or $F_A^-=0$ depending on 
the sign of the instanton number (the second Chern class)
\[
k= \langle c_2(P), [X]\rangle = \dfrac{1}{8\pi^2} \int_X (|F_A^+|^2-|F^-_A|^2) dvol_X.
\]
When $k<0$, connections satisfying
\begin{equation}\label{eq:ASD}
F_A^+ = 0
\end{equation}
are called \emph{anti-self-dual} (ASD) connections, which are special cases of Yang-Mills connections due to the Bianchi identity.

Thus the Yang--Mills functional admits absolute minima in each topological
sector, attained precisely by ASD connections.
This reduction from a second-order variational problem to a first-order
elliptic equation is the four-dimensional analogue of the Bogomolny reduction
for monopoles in three dimensions.

\subsubsection*{Ellipticity and moduli spaces}

The ASD equation \eqref{eq:ASD} is invariant under gauge transformations.
To study its solutions, one considers the moduli space
\[
\mathcal{M}(P)
:= \{ A \mid F_A^+=0 \}/\mathcal{G},
\]
where $\mathcal{G}$ is the group of gauge transformations.

After imposing a gauge-fixing condition (such as Coulomb gauge), the linearised
ASD operator becomes elliptic.
Under suitable topological assumptions and a generic Riemannian metric, $\mathcal{M}(P)$ is a finite-dimensional
manifold away from singular points corresponding to reducible connections.
Its expected dimension is given by an index of the
Atiyah--Hitchin--Singer deformation complex  \cite{AHS} at an ASD connection $A$
\[
0 \longrightarrow \Omega^0(X,\mathfrak{g}_P)
\xrightarrow{\,d_A\,}
\Omega^1(X,\mathfrak{g}_P)
\xrightarrow{\,d_A^+\,}
\Omega^2_+(X,\mathfrak{g}_P)
\longrightarrow 0,
\]
with $\fg_P = P\times_G \fg$, whose cohomology describes infinitesimal gauge transformations, infinitesimal
deformations, and obstructions, respectively. By the Atiyah-Singer index theorem,  the expected dimension is 
\[
\dim \mathcal{M}(P)
= 8|k| - 3(1-b_1(X)+b_2^+(X)),
\]
where $k$ is the instanton number (second Chern class) and $b_i(X)$ are Betti
numbers of $X$.

These moduli spaces encode subtle global information about the smooth structure
of $X$.

\subsubsection*{Donaldson invariants}

Donaldson’s key insight was that the topology of $\mathcal{M}(P)$ could be used
to define invariants of smooth four-manifolds.
By integrating suitable cohomology classes over the moduli space (or its
compactifications), one obtains numerical invariants that depend on the smooth
structure of $X$.

These Donaldson invariants demonstrated, for the first time, that smooth
four-manifolds exhibit phenomena invisible to homotopy or homeomorphism
classification.
In particular, they showed that many topological four-manifolds admit no
smooth structure compatible with certain intersection forms, and that
$\mathbb{R}^4$ admits exotic smooth structures.

Gauge theory thus entered topology not as an auxiliary tool, but as a primary
source of new invariants.

\subsubsection*{Floer theory, Chern--Simons functional, and categorification}

The passage from four-dimensional gauge theory to three-manifold topology is
mediated by the Chern--Simons functional.
Let $Y$ be a closed, oriented three-manifold, and let $P\to Y$ be a principal
$G$-bundle, with $G=SU(2)$ or $SO(3)$.
On the space $\mathcal{A}(P)$ of connections, the Chern--Simons functional is
defined (up to an additive constant) by
\[
\mathrm{CS}(A)
= \frac{1}{8\pi^2}\int_Y \mathrm{tr}
\left(
A\wedge dA + \frac{2}{3}A\wedge A\wedge A
\right).
\]
Although $\mathrm{CS}$ is not gauge-invariant as a real-valued function, its
exponential $\exp(2\pi i\,\mathrm{CS})$ is gauge-invariant, and its critical
points are precisely the flat connections on $P$.

The formal gradient flow equation of $\mathrm{CS}$, with respect to the
$L^2$-metric on $\mathcal{A}(P)$, is
\begin{equation}\label{eq:CS_gradient_flow}
\frac{dA}{dt} = - * F_A .
\end{equation}
When interpreted on the four-manifold $\mathbb{R}\times Y$, equation
\eqref{eq:CS_gradient_flow} is equivalent to the anti-self-dual Yang--Mills
equation for a connection on the product bundle over $\mathbb{R}\times Y$.
Thus, instantons on a four-manifold with cylindrical ends appear naturally as
gradient trajectories of the Chern--Simons functional.

Floer’s fundamental insight \cite{Floer} was to treat $\mathrm{CS}$ as an infinite-dimensional
Morse function.
The chain complex is generated by (irreducible) flat connections on $Y$, graded
by a relative index determined by spectral flow.
The differential counts finite-energy solutions of the ASD equations on
$\mathbb{R}\times Y$ connecting critical points.
The resulting homology groups, denoted $HF_*(Y)$, are invariants of the
three-manifold.

Analytically, this construction requires overcoming substantial difficulties:
the configuration space is infinite-dimensional and singular, the functional
is only Morse--Bott in general, and bubbling phenomena must be controlled.
Nevertheless, Floer’s theory established a rigorous framework in which
three-manifold topology could be studied using four-dimensional gauge theory.

For an integral homology three-sphere $Y$, the Casson invariant $\lambda(Y)$
counts (with signs) certain Lagrangian intersections associated to a Heeggard splitting of $Y$.
In a seminal result, Taubes \cite{Taubes} showed that the Casson invariant can be interpreted
in terms of gauge theory, identifying it with a suitably defined count of
flat $SU(2)$-connections on $Y$.

Floer homology refines this picture categorically.
For a homology sphere $Y$, the Euler characteristic of instanton Floer homology
recovers the Casson invariant:
\[
\chi\bigl(HF_*(Y)\bigr) = 2\,\lambda(Y).
\]
In this precise sense, Floer homology provides a \emph{categorification} of the
Casson invariant and of Taubes’ gauge-theoretic interpretation: instead of a
single numerical invariant, one obtains a graded homology theory whose
alternating sum reproduces the classical count.

This categorification marks a decisive conceptual advance.
It transforms a counting invariant into a homological structure, making
visible additional layers of topological information and functoriality.
Moreover, it establishes a deep link between three- and four-dimensional gauge
theory: three-manifold invariants arise as shadows of four-dimensional
instanton moduli spaces.

From a broader perspective, Floer theory initiated a paradigm that has since
permeated geometry and topology.
Analogous constructions now appear in symplectic geometry, low-dimensional
topology, and mirror symmetry, all tracing their lineage to the gauge-theoretic
insights originating in Yang--Mills theory.

%------------------------------------------------
\subsection{Reduction to surfaces and the Atiyah--Floer conjecture}
\label{subsec:Atiyah_Floer}
%------------------------------------------------

A complementary and conceptually illuminating reduction of four-dimensional
Yang--Mills theory arises when one considers three-manifolds that are fibred,
or decomposed, over a two-dimensional Riemann surface.
This perspective reveals a deep and still only partially understood
relationship between gauge theory and symplectic geometry, encapsulated in the
Atiyah--Floer conjecture \cite{Atiyah88}.

\subsubsection*{Reduction along a surface}

Let $Y$ be a closed three-manifold that admits a Heegaard splitting
\[
Y = H_1 \cup_\Sigma H_2,
\]
where $\Sigma$ is a closed, oriented Riemann surface and $H_1,H_2$ are handlebodies.
Gauge-theoretically, one may study flat $G$-connections on $Y$ by restricting
them to $\Sigma$.
This leads naturally to the moduli space
\[
\mathcal{M}_\Sigma := \{ \text{flat } G\text{-connections on } \Sigma \}/\mathcal{G},
\]
which, away from singularities, is a finite-dimensional symplectic manifold.
The symplectic structure arises from the Atiyah--Bott construction, identifying
$\mathcal{M}_\Sigma$ as an infinite-dimensional symplectic quotient.

Each handlebody $H_i$ determines a Lagrangian submanifold
\[
L_i \subset \mathcal{M}_\Sigma,
\]
consisting of flat connections on $\Sigma$ that extend over $H_i$.
Thus, from the two-dimensional viewpoint, the topology of $Y$ is encoded by a
pair of Lagrangian submanifolds inside a symplectic moduli space.

\subsubsection*{Gauge-theoretic versus symplectic Floer theories}

On the gauge-theoretic side, instanton Floer homology associates to $Y$ a
homology group defined by counting anti-self-dual connections on
$\mathbb{R}\times Y$.
On the symplectic side, one may attempt to define a Lagrangian Floer homology
group as in \cite{FOOO}
\[
HF(L_1,L_2)
\]
by counting pseudo-holomorphic strips in $\mathcal{M}_\Sigma$ with boundary on
$L_1$ and $L_2$.

Formally, both theories are Floer homologies: one infinite-dimensional and
elliptic modulo gauge, the other finite-dimensional but analytically delicate
due to singularities and bubbling.
The Atiyah--Floer conjecture asserts that these two constructions are, under
appropriate hypotheses, equivalent.

\subsubsection*{Statement of the Atiyah--Floer conjecture}

In its original formulation, the Atiyah--Floer conjecture \cite{Atiyah88} predicts an
isomorphism
\[
HF_{\mathrm{inst}}(Y)
\;\cong\;
HF_{\mathrm{Lag}}(L_1,L_2),
\]
relating instanton Floer homology of the three-manifold $Y$ to the Lagrangian
Floer homology of the corresponding pair of Lagrangians in the moduli space of
flat connections on $\Sigma$.

Conceptually, the conjecture asserts that:
\begin{itemize}
\item four-dimensional gauge theory on $\mathbb{R}\times Y$,
\item three-dimensional Chern--Simons theory on $Y$,
\item and two-dimensional symplectic geometry on $\Sigma$,
\end{itemize}
are different manifestations of a single underlying structure.
It may be viewed as a precise mathematical expression of the idea that gauge
theory admits a ``dimensional reduction'' to symplectic topology.
 
Despite its conceptual clarity, the Atiyah--Floer conjecture is notoriously
difficult.
The moduli space $\mathcal{M}_\Sigma$ is typically singular, noncompact, and
fails to be monotone.
On the gauge-theoretic side, reducible connections and bubbling phenomena
complicate transversality.
Establishing a direct correspondence between ASD instantons and
pseudo-holomorphic curves requires controlling adiabatic limits, boundary
conditions, and compactness simultaneously.
These difficulties explain why the conjecture has resisted a complete proof
for several decades, while nevertheless inspiring vast developments on both
sides of the gauge–symplectic divide.

In recent years, significant progress has been made toward a rigorous
formulation of the Atiyah--Floer correspondence.
Notably, work of Daemi, Fukaya and Lipyanskiy \cite{DFL}  has introduced new analytical frameworks that
bridge instanton Floer theory and Lagrangian Floer theory via higher
categorical and $A_\infty$-theoretic structures.
While a complete proof of the Atiyah--Floer conjecture in full generality
remains open, these developments strongly suggest that the conjecture is
correct at a structural level.
They also clarify the precise sense in which gauge theory in four dimensions
controls symplectic topology in two dimensions.

The Atiyah--Floer conjecture occupies a central position in the applications of gauge theory to low dimensional topology.
It provides a conceptual bridge between Donaldson--Floer theory, symplectic
geometry, and, ultimately, mirror symmetry.
It exemplifies how ideas originating in
Yang--Mills theory continue to generate deep connections across dimensions and
disciplines, reinforcing the unifying legacy of gauge theory in mathematics.
%------------------------------------------------
\subsection{Dimensional reductions of ASD Yang--Mills: monopoles and Hitchin systems,
and their roles in 2D/3D mirror symmetry}
\label{subsec:reductions_monopoles_hitchin}
%------------------------------------------------

A striking feature of four-dimensional Yang--Mills theory is that several of
its most important geometric structures in lower dimensions arise as
\emph{dimensional reductions} of the anti-self-dual (ASD) equation
\[
F_A^+ = 0
\qquad \text{on a 4-manifold.}
\]
Two reductions are particularly foundational: the reduction to magnetic
monopoles in three dimensions and the reduction to Hitchin’s equations on a
Riemann surface.
Both yield elliptic moduli problems with hyperkähler geometry and both
reappear naturally in modern duality and mirror symmetry.

%------------------------------------------------
\subsubsection*{Reduction to three-dimensional magnetic monopoles (Bogomolny equations)}
%------------------------------------------------

Consider the ASD equations on $\mathbb{R}^4$ with coordinates
$(x^1,x^2,x^3,x^4)$, and impose translation-invariance in the $x^4$-direction.
Write the connection on $\mathbb{R}^4$ as
\[
\mathbb{A} = A + \Phi\,dx^4,
\]
where $A=\sum_{i=1}^3 A_i\,dx^i$ is a connection on $\mathbb{R}^3$ and
$\Phi$ is an adjoint-valued scalar field (interpreted as the $x^4$-component of
the 4D connection).  Assume $\partial_{x^4}A=\partial_{x^4}\Phi=0$.

A direct curvature computation gives
\[
F_{\mathbb{A}}
= F_A + (D_A\Phi)\wedge dx^4,
\]
where $F_A$ is the curvature of $A$ on $\mathbb{R}^3$ and $D_A\Phi=d\Phi+[A,\Phi]$.
The ASD equation $F_{\mathbb{A}} + *_{\mathbb{R}^4} F_{\mathbb{A}}=0$
reduces precisely to the \emph{Bogomolny monopole equation} (\cite{Bogo1976, Murray1984}) on  $\mathbb{R}^3$:
\begin{equation}\label{eq:Bogomolny_reduction}
F_A = *_{\mathbb{R}^3} D_A\Phi.
\end{equation}

Thus, 3D monopoles are not an additional structure appended to gauge theory;
they are literally the shadow of 4D self-duality under one-dimensional
reduction.  The moduli space of framed solutions of \eqref{eq:Bogomolny_reduction}
(with finite-energy boundary conditions, typically enforced by symmetry
breaking) inherits a natural hyperkähler metric, and in low charges it admits
explicit and remarkable geometry (e.g.\ the Atiyah--Hitchin metric in charge
$2$ in \cite{AtiyahHitchin}).

This reduction is also the conceptual gateway to the \emph{Nahm transform} in \cite{Nahm}:
further symmetry assumptions (or alternative reductions) lead from the monopole
equations to the Nahm equations, revealing a tight link between monopoles,
integrable ODE, and algebraic geometry.

%------------------------------------------------
\subsubsection*{Reduction to Hitchin’s self-duality equations on a Riemann surface}
%------------------------------------------------

A second fundamental reduction arises by considering ASD Yang--Mills on a
product of two Riemann surfaces.
Let $C$ be a compact Riemann surface with local complex coordinate $z$.
Consider ASD connections on $X=C\times \mathbb{R}^2$ (or $C\times T^2$), and
impose translation-invariance along the $\mathbb{R}^2$-directions.
Equivalently, write a connection on $C\times \mathbb{R}^2$ as
\[
\mathbb{A} = A + \phi_1\,dx^1 + \phi_2\,dx^2,
\]
where $A$ is a connection on $C$ and $\phi_1,\phi_2$ are adjoint-valued scalars.
Introduce the complex Higgs field
\[
\varphi := (\phi_1 - i\phi_2)\,dz \in \Omega^{1,0}(C,\mathrm{ad}P),
\]
and write $A=A^{1,0}+A^{0,1}$ with $\bar\partial_A$ the induced $(0,1)$-operator.

Then the ASD equations reduce to \emph{Hitchin’s equations} \cite{Hitchin87} on $C$:
\begin{align}
F_A + [\varphi,\varphi^\ast] &= 0, \label{eq:Hitchin1}\\
\bar\partial_A \varphi &= 0. \label{eq:Hitchin2}
\end{align}
Here $\varphi^\ast$ is the adjoint with respect to a chosen Hermitian metric,
and \eqref{eq:Hitchin2} says that $\varphi$ is holomorphic with respect to the
holomorphic structure defined by $A^{0,1}$.

The moduli space $\mathcal{M}_{\mathrm{H}}(C,G)$ of solutions
$(A,\varphi)$ modulo gauge transformations is a hyperkähler manifold (with
singularities in general), admitting:
\begin{itemize}
\item a complex structure in which points correspond to stable Higgs bundles;
\item a complex structure in which points correspond to flat $G_{\mathbb{C}}$-connections;
\item the \emph{Hitchin fibration}, an algebraically completely integrable system
whose generic fibres are abelian varieties.
\end{itemize}
Thus Hitchin’s equations provide a canonical meeting point of gauge theory,
complex geometry, and integrable systems.

%------------------------------------------------
\subsubsection*{Appearance in 2D mirror symmetry \cite{HT2003}}
%------------------------------------------------

Hitchin moduli spaces are central examples in modern 2D mirror symmetry.
In one direction, $\mathcal{M}_{\mathrm{H}}(C,G)$ is a hyperkähler manifold
and hence supports families of 2D $\mathcal{N}=(4,4)$ sigma models.
Mirror symmetry for such theories is naturally formulated as a hyperkähler
version of SYZ  mirror conjecture\cite{SYZ}: mirror partners are expected to arise by dualising the
(typically torus) fibres of an integrable system.

The Hitchin fibration furnishes precisely such a structure.
Roughly speaking, for Langlands-dual groups $G$ and $^LG$, the Hitchin systems
for $G$ and $^LG$ have dual abelian fibres over a common base, and this duality
underlies the expectation that the corresponding sigma models are mirror.
In this way, Hitchin’s reduction of ASD Yang--Mills provides one of the most
geometric realisations of mirror symmetry in a Calabi--Yau setting, with
branes in $\mathcal{M}_{\mathrm{H}}$ playing the role of the mirror objects.

%------------------------------------------------
\subsubsection*{Appearance in 3D mirror symmetry (monopole moduli and Higgs/Coulomb branches \cite{IS96, HW97,Gaiotto08,BFN18})}
%------------------------------------------------

In three-dimensional $\mathcal{N}=4$ supersymmetric gauge theories, the
geometry of moduli spaces is governed by two hyperkähler manifolds: the
\emph{Higgs branch} and the \emph{Coulomb branch}.
A hallmark of 3D mirror symmetry is that it exchanges these branches between a
pair of mirror-dual theories.

From the gauge-theoretic viewpoint, the Coulomb branch is closely tied to
monopole configurations and their moduli: monopole operators, and the
associated moduli problems, encode the nonperturbative geometry of the Coulomb
branch.
This makes the reduction ASD $\Rightarrow$ Bogomolny monopoles
\eqref{eq:Bogomolny_reduction} conceptually fundamental for 3D mirror symmetry.

At the same time, Higgs-branch geometry is typically described by hyperkähler
quotients (moment-map equations), which are themselves gauge-theoretic in
origin.
Thus, monopoles and Hitchin-type systems may be viewed as geometric avatars of
the two sides of 3D mirror symmetry: monopole moduli naturally encode Coulomb
data, while Higgs-bundle-type moduli encode Higgs data, and mirror symmetry
interchanges these hyperkähler geometries.

Dimensional reductions of ASD Yang--Mills provide a unified explanation for why
monopole moduli spaces and Hitchin moduli spaces appear ubiquitously in
geometry and physics.
They arise as canonical lower-dimensional shadows of four-dimensional
self-duality, and their hyperkähler geometry makes them natural arenas for
both 2D and 3D mirror symmetry.
In this sense, mirror symmetry does not merely coexist with gauge theory; it
is repeatedly \emph{generated} by the geometric structures that gauge theory
creates across dimensions.

%================================================
\section{From Yang to Baxter: Yang--Baxter, Chern--Simons, and Quantum Groups}
\label{sec:YB_quantum_groups}
%================================================

\noindent\textbf{Transition from gauge theory to integrability.}
The gauge-theoretic narrative of Sections~\ref{subsec:ASD_Donaldson}--\ref{subsec:reductions_monopoles_hitchin}
reveals a recurring theme: the Yang--Mills equations generate rich moduli spaces,
and these moduli problems often carry hidden algebraic structures (symplectic,
hyperk\"ahler, and integrable).
A second, equally consequential strand of Yang’s legacy emerges from a different
question: when does an interacting many-body system admit \emph{exact}
factorisation and solvability?
The answer, in both quantum mechanics and statistical mechanics, is governed by
the same algebraic consistency constraint—the Yang--Baxter equation.
In the modern viewpoint, this constraint is not isolated: it reappears
geometrically in Chern--Simons theory, where Wilson lines braid, and algebraically
in quantum groups, which provide canonical families of $R$-matrices.
Line and boundary defects in three-dimensional topological gauge theories make
these $R$-matrices visible as braiding operators, thereby connecting gauge theory
to quantum integrability in a direct and structural way. \cite{Witten1989}

%------------------------------------------------
\subsection{Yang: factorised scattering and consistency}
\label{subsec:yang_factorised_scattering}
%------------------------------------------------

The earliest appearance of what is now called the Yang--Baxter equation arose in
the work of Yang
on one-dimensional quantum many-body systems with $\delta$-function interactions
\cite{Yang1967,Yang1968}.
In these papers, Yang was not seeking an abstract algebraic structure; rather,
he was analysing the precise conditions under which an interacting many-body
quantum system admits an exact solution.

\subsubsection*{One-dimensional scattering and rigidity}

In one spatial dimension, particles cannot pass each other without interacting.
As a consequence, the scattering of $N$ particles is highly constrained.
Yang observed that in certain special models the full $N$-body scattering matrix
factorises into a product of two-body scattering matrices.
This factorisation is not an approximation but an exact property of the model.

Let $V$ denote the internal state space of a single particle.
For particles labelled by $i,j$, let
\[
S_{ij} \in \mathrm{End}(V^{\otimes N})
\]
denote the two-body scattering operator acting nontrivially only on the $i$-th
and $j$-th tensor factors.
Factorised scattering asserts that the full scattering operator can be written
as an ordered product of such $S_{ij}$’s corresponding to successive pairwise
collisions.

\subsubsection*{Consistency for three particles}

For $N \ge 3$, the same physical scattering process admits multiple decompositions
into two-body scatterings, corresponding to different orderings of collisions.
Yang’s key insight was that physical consistency demands that all such
factorisations yield the same total scattering operator.

For three particles, this requirement leads to the fundamental condition
\begin{equation}\label{eq:YBE_Yang_original}
S_{12} S_{13} S_{23} = S_{23} S_{13} S_{12}
\quad \text{on } V^{\otimes 3}.
\end{equation}
This equation appears explicitly in Yang’s analysis as the necessary and
sufficient condition for factorised scattering to be well defined.

Up to conventional placement of permutation operators (which distinguish
between $S$-matrices and $R$-matrices in later formulations),
\eqref{eq:YBE_Yang_original} is precisely the quantum Yang--Baxter equation.

What is historically decisive in Yang’s work is that the equation
\eqref{eq:YBE_Yang_original} emerges as a \emph{consistency condition},
not as an algebraic postulate.
Local two-body interactions determine global many-body dynamics if and only if
this identity holds.
In this sense, the Yang--Baxter equation is born as a principle of coherence:
different ways of assembling local processes must agree.

Yang himself did not pursue the abstract algebraic consequences of this
equation.
Nevertheless, his formulation provided the first clear instance in which the
Yang--Baxter equation appears as a fundamental constraint dictated by quantum
mechanics.

%------------------------------------------------
\subsection{Baxter: solvable lattice models and commuting transfer matrices}
\label{subsec:baxter_lattice_models}
%------------------------------------------------

Independently of Yang’s work, a parallel consistency principle emerged in
statistical mechanics through the work of
Rodney Baxter,
culminating in his systematic theory of exactly solvable lattice models
\cite{Baxter72,Baxter82}.
Baxter’s approach addressed a different problem: how to compute thermodynamic
quantities exactly in strongly interacting two-dimensional systems.

\subsubsection*{Transfer matrices and exact solvability}

In a two-dimensional lattice model, the central object is the transfer matrix
$T(u)$, depending on a spectral parameter $u$, which encodes the Boltzmann weights
of a row (or column) of the lattice.
The partition function of the model is expressed in terms of powers of $T(u)$,
and exact solvability is achieved if the spectral problem of $T(u)$ can be solved
explicitly.

Baxter identified the defining criterion of solvability as the existence of a
commuting family of transfer matrices:
\[
[T(u), T(v)] = 0 \qquad \text{for all } u,v.
\]
This commutativity allows simultaneous diagonalisation and leads to functional
relations governing the spectrum.

\subsubsection*{Local relations and the star--triangle equation}

The commutativity of transfer matrices is guaranteed by a local identity among
Boltzmann weights, known classically as the star--triangle relation.
Baxter showed that this local relation is the fundamental algebraic mechanism
underlying solvability.

In modern language, the star--triangle relation is equivalent to the
Yang--Baxter equation for an $R$-matrix
\[
R(u) \in \mathrm{End}(V \otimes V),
\]
namely
\begin{equation}\label{eq:YBE_Baxter}
R_{12}(u-v)\, R_{13}(u-w)\, R_{23}(v-w)
=
R_{23}(v-w)\, R_{13}(u-w)\, R_{12}(u-v),
\end{equation}
an identity in $\mathrm{End}(V^{\otimes 3})$.
Baxter did not merely recognise this equation; he used it as a constructive
tool.

Baxter’s decisive contribution was to elevate the Yang--Baxter equation from a
consistency condition to a systematic framework.
Starting from an $R$-matrix satisfying \eqref{eq:YBE_Baxter}, one constructs:
\begin{itemize}
\item monodromy matrices,
\item commuting transfer matrices,
\item functional relations and Bethe-type equations,
\end{itemize}
from which exact physical quantities can be derived.

Thus, while Yang encountered the Yang--Baxter equation as a constraint emerging
from quantum scattering, Baxter transformed it into the organising principle of
integrable statistical mechanics.

The intellectual unity between Yang and Baxter lies in their shared insistence
on consistency.
For Yang, consistency of multi-particle scattering enforces factorisation.
For Baxter, consistency of local Boltzmann weights enforces commutativity and
exact solvability.
In both cases, local rules determine global structure, and the Yang--Baxter
equation is the algebraic expression of that principle.

Baxter’s construction reveals a crucial shift in perspective.
Once the Yang--Baxter equation is understood as the local condition guaranteeing
the commutativity of transfer matrices, it becomes clear that the equation
controls not merely solvability, but the consistent reordering of interactions.
In this sense, the Yang--Baxter equation encodes the algebraic rules for
exchanging neighbouring degrees of freedom.
Abstracting away from the specific lattice model, these exchange rules acquire
a topological meaning: they define representations of the braid group.
The passage from solvable models to braiding is therefore not an additional
structure, but the natural categorical and topological reformulation of the
same consistency principle.

%------------------------------------------------
\subsection{Braids and quantum groups: Drinfeld and Jimbo}
\label{subsec:braids_quantum_groups}
%------------------------------------------------

By the early 1980s, the Yang--Baxter equation had appeared independently in
one-dimensional scattering theory and in exactly solvable lattice models.
The next decisive step was to recognise that this equation defines an abstract
algebraic structure governing the consistent exchange of degrees of freedom.
This insight emerged through the pioneering work of
 Drinfeld and  Jimbo
who independently introduced what are now called \emph{quantum groups}
\cite{Drinfeld,Jimbo}.

Let $R\in \mathrm{End}(V\otimes V)$ be an invertible solution of the
Yang--Baxter equation
\[
R_{12} R_{13} R_{23} = R_{23} R_{13} R_{12}
\quad \text{in } \mathrm{End}(V^{\otimes 3}).
\]
Define $\check R := P\circ R$, where $P$ is the permutation operator on
$V\otimes V$.
Then the operators
\[
\sigma_i := \check R_{i,i+1}
\]
satisfy the braid relations
\[
\sigma_i \sigma_{i+1} \sigma_i
=
\sigma_{i+1} \sigma_i \sigma_{i+1},
\qquad
\sigma_i \sigma_j = \sigma_j \sigma_i
\ \ (|i-j|\ge 2),
\]
and hence define a representation of the braid group $B_n$ on $V^{\otimes n}$.

This observation, implicit in earlier work, made explicit the topological
content of the Yang--Baxter equation: it governs the consistent exchange of
labels along braids.
The equation thus acquires an interpretation as a coherence condition for
braiding, independent of any particular physical model.

\subsubsection*{Drinfeld--Jimbo quantum groups}

Drinfeld and Jimbo independently discovered that solutions of the Yang--Baxter
equation arise systematically from certain Hopf algebra deformations of
universal enveloping algebras of semisimple Lie algebras.
For a complex semisimple Lie algebra $\mathfrak g$, they introduced a
$q$-deformation $U_q(\mathfrak g)$, generated by elements
$E_i,F_i,K_i^{\pm1}$ subject to $q$-deformed relations.

Drinfeld identified the essential structural feature as
\emph{quasi-triangularity}: the existence of a universal $R$-matrix
\[
\mathcal R \in U_q(\mathfrak g)\widehat\otimes U_q(\mathfrak g)
\]
satisfying
\[
\mathcal R\, \Delta(x)\, \mathcal R^{-1} = \Delta^{\mathrm{op}}(x),
\]
together with compatibility identities implying the universal Yang--Baxter
equation.
Upon evaluation in representations, $\mathcal R$ produces concrete
$R$-matrices solving the Yang--Baxter equation.
Jimbo’s work emphasised the representation-theoretic consequences, constructing
explicit $R$-matrices associated with affine Lie algebras and integrable models.
Together, these works established quantum groups as the natural algebraic
framework underlying Yang--Baxter structures.

What distinguishes the Drinfeld--Jimbo construction is that the Yang--Baxter
equation is no longer an isolated identity.
It is encoded in the axioms of a Hopf algebra deformation, and braid group
actions arise functorially on tensor products of representations.
Thus, the consistency principle first encountered by Yang and Baxter becomes
internalised as algebraic structure.
In this way, quantum groups mediate between integrable models, representation
theory, and topology, preparing the ground for a geometric interpretation of
braiding.
%------------------------------------------------
\subsection{Chern--Simons gauge theory and the Jones polynomial: Witten's synthesis}
\label{subsec:chern_simons_witten}
%------------------------------------------------

The decisive geometric synthesis of the Yang--Baxter equation, braid groups,
and knot invariants was achieved by
 Witten
in his seminal 1989 paper
\emph{Quantum Field Theory and the Jones Polynomial}
\cite{Witten1989}.
In this work, Witten showed that the Jones polynomial and its generalisations
arise naturally from the quantisation of three-dimensional Chern--Simons gauge
theory.
This result provided a profound bridge between quantum gauge theory, topology,
and the algebraic structures discovered by Jones, Drinfeld, and Jimbo.

\subsubsection*{Chern--Simons theory as a topological gauge theory}

Let $Y$ be a closed oriented three-manifold and let $G$ be a compact, simple Lie
group.
For a connection $A$ on a principal $G$-bundle over $Y$, the Chern--Simons
functional is defined by
\begin{equation}\label{eq:CS_action}
S_{\mathrm{CS}}(A)
=
\frac{k}{4\pi}
\int_Y
\mathrm{Tr}\!\left(
A \wedge dA + \frac{2}{3} A \wedge A \wedge A
\right),
\end{equation}
where $k\in\mathbb{Z}$ is the level.
Although \eqref{eq:CS_action} is not strictly gauge-invariant, its exponential
$\exp(i S_{\mathrm{CS}})$ is invariant under all gauge transformations provided
$k$ is integral.

A central feature of Chern--Simons theory is that it is independent of any
choice of metric on $Y$.
Its classical equations of motion,
\[
F_A = 0,
\]
assert that critical points are flat connections.
Thus, the theory is topological in nature, with observables depending only on
the topology of $Y$ and embedded submanifolds.

\subsubsection*{Wilson loops and knot observables}

Given an oriented knot or link $L \subset Y$ and a finite-dimensional
representation $V$ of $G$, Witten introduced Wilson loop observables
\begin{equation}\label{eq:Wilson_loop}
W_V(L)
=
\mathrm{Tr}_V\,\mathrm{P}\exp\!\left(\int_L A\right),
\end{equation}
where $\mathrm{P}$ denotes path ordering.
Formally, the expectation value
\[
\langle W_V(L) \rangle
=
\int \mathcal{D}A\; e^{i S_{\mathrm{CS}}(A)}\, W_V(L)
\]
defines a topological invariant of the pair $(Y,L)$.

Specialising to $Y=S^3$, $G=SU(2)$, and $V$ the fundamental representation,
Witten showed that $\langle W_V(L) \rangle$ coincides with the Jones polynomial
of $L$, evaluated at a root of unity determined by the level $k$.
More generally, other choices of $G$ and $V$ yield the HOMFLY--PT and Kauffman
polynomials.

\subsubsection*{Braiding, monodromy, and the Yang--Baxter equation}

A key insight of Witten’s construction is that knot invariance arises from the
topology of Wilson line worldlines in three dimensions.
When two Wilson lines are exchanged, the corresponding operators undergo a
nontrivial transformation on the quantum theory.
Consistency of multiple exchanges requires that these transformations satisfy the
braid relations.

In the operator formalism, this braiding is governed by an $R$-matrix acting on
tensor products of representations.
The associativity of braiding for three Wilson lines is encoded precisely by
the Yang--Baxter equation.
Thus, the Yang--Baxter relation appears not as an abstract algebraic identity,
but as the condition ensuring topological invariance of Wilson line correlators
under isotopy.

\subsubsection*{Relation to quantum groups}

Witten further observed that the algebraic structure controlling these braidings
is the representation theory of quantum groups at roots of unity.
The $R$-matrix governing the exchange of Wilson lines coincides with the
$R$-matrix obtained from the Drinfeld--Jimbo quantum group $U_q(\mathfrak g)$,
with
\[
q = \exp\!\left(\frac{2\pi i}{k+h^\vee}\right),
\]
where $h^\vee$ is the dual Coxeter number of $\mathfrak g$.

From this perspective, quantum groups arise as effective symmetry algebras of
Chern--Simons theory.
The Jones polynomial is no longer an isolated combinatorial invariant, but a
manifestation of quantum gauge symmetry.

\subsubsection*{Canonical quantisation and conformal field theory}

Witten also analysed Chern--Simons theory via canonical quantisation.
When $Y=\Sigma\times\mathbb{R}$, with $\Sigma$ a Riemann surface, the Hilbert
space of the theory is identified with the space of conformal blocks of the
associated Wess--Zumino--Witten (WZW) model on $\Sigma$.
In this framework, Wilson lines correspond to primary fields, and braiding is
realised through the monodromy of conformal blocks.

This connection explains the deep relationship between knot invariants,
two-dimensional conformal field theory, and affine Lie algebras, and provides
another route through which the Yang--Baxter equation enters the theory.

Witten’s work completed the transition from algebra to geometry.
What had appeared in Yang’s work as a consistency condition for scattering, and
in Baxter’s work as the criterion for exact solvability, was revealed to be a
topological statement about the braiding of worldlines in three dimensions.
Chern--Simons theory thus provides a geometric origin for the Jones polynomial
and for quantum group symmetries.  Witten’s interpretation of the Jones polynomial through Chern--Simons gauge
theory marks a conceptual turning point.
The Yang--Baxter equation, first encountered as a consistency condition in
scattering theory and later as an algebraic identity in integrable models, is
here realised as a topological constraint governing the braiding of Wilson
lines.
In this formulation, algebraic objects such as $R$-matrices and quantum groups
are no longer auxiliary constructions, but effective symmetries emerging from
gauge theory itself.
Once this geometric origin is recognised, it becomes natural to ask how
Yang--Baxter structures persist beyond three-dimensional topology—how they
reappear in higher-dimensional gauge theory, enumerative geometry, and string
theory.
The developments that follow may be viewed as answers to this question: they
trace the continued life of Yang’s and Baxter’s consistency principles in
modern mathematics.

%================================================
\section{Modern Developments: Yang--Baxter Structures from Geometry to Quantum Field Theory}
\label{sec:modern}
%================================================

The geometric reinterpretation of the Yang--Baxter equation initiated by
Chern--Simons theory did not mark an endpoint, but rather the beginning of a new
phase.
In modern mathematical physics, Yang--Baxter structures reappear in unexpected
settings: in the geometry of magnetic monopoles, in quantum cohomology and
enumerative geometry, and ultimately in the foundational questions surrounding
four-dimensional quantum gauge theory.
We review three representative developments that illustrate this evolution.

%------------------------------------------------
\subsection{Atiyah: magnetic monopoles and the Yang--Baxter equation}
\label{subsec:atiyah_monopoles_ybe}
%------------------------------------------------

A striking geometric bridge between Yang--Mills theory and the Yang--Baxter
equation was proposed by
Atiyah in \cite{Atiyah}.  
Atiyah’s observation is that solutions of the Yang--Baxter equation may be
encoded geometrically in the moduli space of magnetic monopoles.

Consider $SU(2)$ Bogomolny monopoles on $\mathbb{R}^3$,
\[
F_A = * D_A \Phi,
\]
with finite energy.
Such monopoles are classified by an integer magnetic charge $k$ and admit a
twistor-theoretic description via spectral curves (\cite{Hitchin83}).
Specifically, to a charge-$k$ monopole one associates an algebraic curve
$C \subset T^*\mathbb{P}^1$, satisfying reality conditions and degree constraints.

Atiyah emphasised that these spectral curves are not arbitrary.
They satisfy additional symmetry and factorisation properties closely resembling
those that arise in integrable systems.

Atiyah proposed that certain families of monopole spectral curves give rise to
solutions of the Yang--Baxter equation.
He interpreted the Yang--Baxter relation as expressing the compatibility of
different ways of composing monopoles, much as it expresses the compatibility of
different factorizations in scattering theory.

In this picture, the Yang--Baxter equation governs the geometry of monopole
moduli spaces rather than algebraic operators.
It becomes a condition on how spectral data glue together, reflecting the
integrability hidden within the Bogomolny equations.

Although exploratory in nature, Atiyah’s work was conceptually inspiring: it
suggested that Yang--Baxter equations arise naturally from classical gauge theory,
without invoking quantum groups or lattice models.
This insight foreshadowed later developments linking gauge theory, integrable
systems, and enumerative geometry.

%------------------------------------------------
\subsection{Maulik--Okounkov: quantum groups and quantum cohomology}
\label{subsec:mo_qg_qh}
%------------------------------------------------

A major advance in the modern understanding of Yang--Baxter structures occurred
with the work of
 Maulik  and
 Okounkov,
who established a deep connection between quantum groups and quantum cohomology
\cite{MO}. 

Let $X$ be a holomorphic symplectic variety, such as a Nakajima quiver variety.
Its equivariant quantum cohomology ring $QH^*_T(X)$ encodes enumerative data of
rational curves in $X$.
Maulik and Okounkov showed that quantum multiplication operators define a flat
connection (the quantum connection) whose monodromy yields representations of
quantum groups.

A central construction in their theory is that of \emph{stable envelopes}, which
depend on a choice of chamber in equivariant parameter space.
Comparing stable envelopes for different chambers produces canonical
$R$-matrices,
\[
R \in \mathrm{End}(H^*_T(X)\otimes H^*_T(X)),
\]
which satisfy the Yang--Baxter equation.
In this way, Yang--Baxter structures arise from purely geometric data.
Quantum groups appear not as abstract algebraic deformations, but as symmetries
governing enumerative geometry.

This work revealed that the Yang--Baxter equation is a universal compatibility
condition for wall-crossing phenomena in geometry.
It controls how enumerative invariants transform under changes of stability
conditions, echoing Yang’s original principle of consistency across different
factorizations.

%------------------------------------------------
\subsection{Quantum Yang--Mills theory and the Clay Millennium problem}
\label{subsec:clay_yang_mills}
%------------------------------------------------

The enduring depth of Yang--Mills theory is perhaps most clearly reflected in
its appearance as one of the Millennium Prize Problems of 
Clay Mathematics Institute \cite{JW}.
While the classical Yang--Mills equations and their moduli spaces have reshaped
geometry and topology, the quantum theory remains mathematically incomplete,
despite its central role in modern physics.

\vspace{2mm}

\noindent{\bf Statement of the Millennium Prize Problem:}
The Clay problem asks for a rigorous construction of four-dimensional quantum
Yang--Mills theory with compact gauge group $G$ (such as $SU(N)$), together with
a proof of the mass gap: the existence of a positive lower bound on the
spectrum of excitations above the vacuum.
\emph{More precisely, one seeks a mathematically well-defined quantum field theory
satisfying the standard axioms (or an accepted equivalent framework), whose
correlation functions and energy spectrum exhibit confinement and mass
generation consistent with physical expectations.}
\vspace{2mm}

From the classical standpoint, Yang--Mills theory is exceptionally rigid.
Its gauge symmetry, variational structure, and elliptic reductions lead to rich
and well-understood moduli problems.
In four dimensions, anti-self-dual solutions yield finite-dimensional moduli
spaces whose topology encodes subtle information about smooth manifolds.
The quantum theory, however, introduces fundamentally new difficulties.
Ultraviolet divergences, nonperturbative effects, and the infinite-dimensional
nature of the configuration space challenge existing mathematical frameworks.
While perturbative renormalisation is well understood, it does not suffice to
define the theory nonperturbatively or to establish the mass gap.

The mass gap problem is deeply geometric in nature.
Physically, it reflects confinement and the absence of long-range gauge
excitations; mathematically, it is tied to the behaviour of the Yang--Mills
measure on the space of connections modulo gauge transformations.
Despite the success of symmetry breaking mechanisms in electroweak theory, pure
Yang--Mills theory is believed to generate a mass scale dynamically, without
scalar fields—a phenomenon that remains beyond current rigorous methods.
Notably, many of the structures discussed earlier in this article—instantons,
monopoles, Chern--Simons theory, and Yang--Baxter integrability, arise as
controlled or lower-dimensional limits of the full quantum theory.
They provide valuable insights, yet a complete synthesis capable of resolving
the four-dimensional quantum problem has not yet emerged.

Taken together, these advances demonstrate that the Yang--Baxter equation and
quantum groups continue to generate new mathematics well beyond their original
contexts.
They now appear as structural constraints governing topology, geometry, and
quantisation across dimensions.
From this vantage point, the contributions of Yang and Baxter are not confined
to a historical chapter but remain active forces shaping contemporary research.

%================================================
\section{Epilogue: From Symmetry to Mathematical Unity}
%================================================

The legacies of Chen-Ning Yang and  Rodney Baxter
define a remarkable arc in modern mathematical physics.
Although their work emerged from different problems—Yang from gauge symmetry,
Baxter from integrable models—it is unified by a shared principle: that local
consistency, when imposed without compromise, determines global structure.

Yang’s introduction of non-abelian gauge theory placed symmetry at the foundation
of fundamental interactions.
The early obstruction posed by mass did not weaken this principle but instead
forced deeper developments, leading to spontaneous symmetry breaking,
renormalisability, and the geometric theory of connections, instantons, and
monopoles.
What began as a physical postulate evolved into a framework that reshaped
geometry itself.

Baxter’s work followed an equally rigorous path.
In exactly solvable models, consistency was not an aesthetic guide but a defining
requirement.
The Yang–Baxter equation, elevated from a constraint to a methodology, revealed
integrability as a rigid and far-reaching mathematical structure.  
Gauge theory and integrability intersect in quantum groups, Chern–Simons theory,
and the geometry of moduli spaces, where the Yang–Baxter equation reappears as a
geometric and topological condition.
The enduring influence of Yang and Baxter lies not only in the theories they
created, but in the standard they set.
Their work reminds us that insisting on coherence, even in the face of
incompleteness, is often what opens the deepest mathematical paths.

%===============================================
\section*{Acknowledgements}
%================================================
This work is completed during my visit to BICMR, Peking University. I would like to thank Professors Gang Tian and  Xiaobo Liu for their hospitality and providing the inspiring research environment.

I met Chen–Ning Yang only once at an international conference in Nankai when
I was a graduate student. 
By contrast, I was fortunate enough to have Rodney Baxter as a colleague for
nearly twenty years, benefiting directly from his insight, generosity, and
clarity of thought.
In many different ways, both have served as enduring mathematical lighthouses for
me, guiding my understanding of what depth, coherence, and elegance can mean in
mathematical physics.

Over the course of my mathematical career, my engagement with these ideas has
been shaped by many conversations with generous and
inspiring teachers, colleagues and collaborators.  I express my deepest gratitude to them all,  far too many to list here exhaustively.

Lastly, I apologise sincerely for any omissions or inaccuracies in this memorial article.
They reflect only my own limitations and ignorance, and not the extraordinary
breadth and depth of the contributions made  by Yang, Baxter, and the many great
minds whose ideas have shaped these fields. 

%================================================


\begin{thebibliography}{99}


\bibitem{Atiyah} M. Atiyah,
     \emph{Magnetic monopoles and the Yang–Baxter equations,}
     Int. J. Mod. Phys. A7 (1992), 1–15.

    \bibitem{Atiyah88}
 M. Atiyah, \emph{New invariants of 3- and 4-dimensional manifolds}, The mathematical heritage of Hermann Weyl (Durham, NC, 1987), 1988, pp. 285–299.
 
 \bibitem{AtiyahHitchin}
M. Atiyah and N. Hitchin,
\emph{The Geometry and Dynamics of Magnetic Monopoles}, 
Princeton Univ.~Press (1988).

\bibitem{AHS} M. Atiyah, N. Hitchin and I. Singer,
\emph{Self-duality in four-dimensional Riemannian geometry}, Proc. R. Soc. Lond. A. 362, 425-461 (1978). 

\bibitem{Baxter17}
Baxter, R. J., \emph{An accidental academic}, http://hdl.handle.net/1885/117792

\bibitem{Baxter72} Baxter, R.J. \emph{Partition function for the eight-vertex lattice model.} Ann. Phys. 1972, 70, 193–228.

\bibitem{Baxter82}
R.~J.~Baxter,
\emph{Exactly Solved Models in Statistical Mechanics},
Academic Press, 1982.


\bibitem{Bogo1976} E. B. Bogomolny, \emph{Stability of classical solutions,}  Sov. J. Nucl. Phys. 24 (1976), 449–454.


\bibitem{BFN18}
A.~Braverman, M.~Finkelberg, and H.~Nakajima.
\emph{Towards a mathematical definition of Coulomb branches of $3$-dimensional $\mathcal{N}=4$ gauge theories, II}.
Adv.\ Theor.\ Math.\ Phys.  22 (2018), 1071--1147.

\bibitem{BroEng1964}
 R. Brout and F. Englert, 
\emph{Broken symmetry and the mass of gauge vector mesons},
Phys. Rev. Lett. \textbf{13} (1964), 321--323.

\bibitem{DFL} A. Daemi, K.  Fukaya and M. Lipyanskiy, \emph{Lagrangians, SO(3)-instantons and the Atiyah-Floer conjecture}, arXiv:2109.07038.

\bibitem{Donaldson}
S.~K.~Donaldson,
\emph{Floer homology groups in Yang-Mills theory},
Cambridge University Press.

\bibitem{Drinfeld}
V.~G.~Drinfeld,
\emph{Quantum groups},
Proc.\ ICM (Berkeley, 1986), 798--820.


 \bibitem{Floer}
A.~Floer,
An instanton-invariant for 3-manifolds,
\emph{Commun.\ Math.\ Phys.} \textbf{118} (1988), 215--240.

\bibitem{FOOO} K. Fukaya, Y. Oh, H. Ohta and K. Ono, \emph{Lagrangian intersection Floer theory: anomaly and obstruction}, AMS/IP Studies in Adv. Math. 46, 1-2, 2009.

\bibitem{Gaiotto08}
D.~Gaiotto.
\textit{Monopole operators and mirror symmetry in three dimensions}.
J.\ High Energy Phys. \textbf{12} (2008), 078.

\bibitem{HW97}
A.~Hanany and E.~Witten.
\emph{Type IIB superstrings, BPS monopoles, and three-dimensional gauge dynamics}.
Nucl.\ Phys.\ B  492 (1997), 152--190.

\bibitem{HT2003} T. Hausel and M. Thaddeus, \emph{Mirror symmetry, Langlands duality, and the Hitchin system}. Invent. Math., 153(1):197–229, 2003.



\bibitem{Higgs1964}
P. W. Higgs,
\emph{Broken symmetries and the masses of gauge bosons},
Phys. Rev. Lett. \textbf{13} (1964).

\bibitem{Hitchin87} N. Hitchin, \emph{The self-dual equations on a Riemann surface}, 
Proc. London Math. Soc. (3) 55 (1987) 59-126.


\bibitem{Hitchin83}
N.~J.~Hitchin,
Monopoles and geodesics,
\emph{Commun.\ Math.\ Phys.} \textbf{83} (1982), 579--602.

\bibitem{tHooft} G. ’t Hooft,
    \emph{Magnetic monopoles in unified gauge theories,}
Nuclear Physics B 79 (1974), 276–284.

\bibitem{IS96}
K.~Intriligator and N.~Seiberg.
\emph{Mirror symmetry in three-dimensional gauge theories}.
Phys.\ Lett.\ B 387 (1996), 513--519.

\bibitem{JW} A. Jaffe and E. Witten, \emph{Quantum Yang-Mills theory},  https://www.claymath.org/wp-content/uploads/2022/06/yangmills.pdf.

    
\bibitem{Jimbo}  M.~Jimbo, \emph{A $q$-difference analogue of $U(\mathfrak{g})$ and the Yang--Baxter equation},
 Lett.\ Math.\ Phys.\ \textbf{10} (1985), 63--69.

\bibitem{MO} D. Maulik and  A. Okounkov, \emph{Quantum groups and quantum cohomology}, arXiv:1211.1287. 

\bibitem{Murray1984} M. K. Murray, \emph{Non-abelian magnetic monopoles}, Commun. Math. Phys. 96,539-565.

\bibitem{Nahm} W. Nahm, \emph{The construction of all self-dual monopoles by the ADHM mnethod}, Monopoles in QFT, Proc. of the monopole conference in Trieste 1981, World Scientific, Singapore, 1986. 

\bibitem{Polyakov}  A. M. Polyakov,
    Particle spectrum in quantum field theory,
    JETP Lett. 20 (1974), 194–195.

\bibitem{SYZ}  A. Strominger, S. T. Yau and E. Zaslow \emph{Mirror symmetry is 
T-duality},  Nuclear Phys. B 479.1-2, 1996, pp. 243–259.

\bibitem{Taubes}  C. Taubes, \emph{Casson's invariants and gauge theory},  JDG., 
31(1990) 547-599. 



 \bibitem{Witten1989}
E.~Witten, \emph{Quantum field theory and the Jones polynomial},
 Comm.\ Math.\ Phys.\ \textbf{121} (1989), 351--399.





\bibitem{YangMills1954}
C.~N.~Yang and R.~L.~Mills,
\emph{Conservation of isotopic spin and isotopic gauge invariance},
Phys.\ Rev.\ \textbf{96} (1954), 191--195.

\bibitem{Yang1967}
C.~N.~Yang,
\emph{Some exact results for the many-body problem in one dimension with repulsive delta-function interaction},
Phys.\ Rev.\ Lett.\ \textbf{19} (1967), 1312--1315. 

\bibitem{Yang1968}
C.~N.~Yang,
\emph{$S$ matrix for the one-dimensional $N$-body problem with repulsive or attractive $\delta$-function interaction},
Phys.\ Rev.\ \textbf{168} (1968), 1920--1923. 

\end{thebibliography}
\end{document}